\documentstyle[12pt]{article}

\title{A NEW IDENTITY IN MINKOWSKI SPACE AND SOME APPLICATIONS OF IT}

\author {Alexander~L.~Bondarev
\and \it National Scientific and Educational Center of Particle and
\and \it High Energy Physics attached to Belarusian State University
\and \it M.Bogdanovich str.,153, Minsk 220040, Republic of Belarus
\and \rm e-mail: bondarev@hep.by}

\begin{document}
\maketitle

Published in \\
{\it Teoreticheskaya i Matematicheskaya Fizika},
Vol.101, No.2, P.315 -- 319 (1994) (in Russian)

Translated in \\
{\it Theoretical and Mathematical Physics},
Vol.101, No.2, P.1376 -- 1379 (1994)  \\

\begin{abstract}
    Schouten's  identity  is  used  to  obtain  a  new identity in
Minkowski  space.    Some  applications  of  the  new  identity in
high-energy physics are  considered, including the  possibility of
significant  shortening  of  the  expressions  for  the  traces of
products of 10 and more Dirac ${\gamma}$ matrices.
\end{abstract}

We write Schouten's identity in the form\footnotemark
\footnotetext[1]
{
We use a metric in which
$
\displaystyle
a_{\mu} = ( \vec{a}, \; a_4 = {\it i} a_0 ) , \;
ab = a_{\mu} b_{\mu} = \vec{a} \vec{b} - a_0 b_0 .
$
}
\begin {equation}
\displaystyle
  {\delta}_{\mu \nu} {\varepsilon}_{\alpha \beta \lambda \rho}
+ {\delta}_{\mu \alpha} {\varepsilon}_{\beta \lambda \rho \nu}
+ {\delta}_{\mu \beta} {\varepsilon}_{\lambda \rho \nu \alpha}
+ {\delta}_{\mu \lambda} {\varepsilon}_{\rho \nu \alpha \beta}
+ {\delta}_{\mu \rho} {\varepsilon}_{\nu \alpha \beta \lambda}
\equiv 0 \;\; .
\label{e1}
\end {equation}

    The validity of this identity  follows from the fact that  the
expression  on  the  left-hand  side  of  (\ref{e1}) is completely
antisymmetric  with  respect  to  each  of five indices:  ${\nu}$,
${\alpha}$, ${\beta}$, ${\lambda}$, ${\rho}$.  In four-dimensional
space, every  tensor that  is antisymmetric  with respect  to more
than four  indices vanishes  identically, since  the values  of at
least two of them must be equal.

We multiply (\ref{e1}) by
${\varepsilon}_{\sigma \tau \kappa \omega}$:
\begin {equation}
\begin {array}{l} \displaystyle
({\delta}_{\mu \nu    }{\varepsilon}_{\alpha \beta \lambda \rho} +
 {\delta}_{\mu \alpha }{\varepsilon}_{\beta \lambda \rho \nu   } +
 {\delta}_{\mu \beta  }{\varepsilon}_{\lambda \rho \nu \alpha  } +
 {\delta}_{\mu \lambda}{\varepsilon}_{\rho \nu \alpha \beta    } +
 {\delta}_{\mu \rho   }{\varepsilon} _{\nu \alpha \beta \lambda})
{\varepsilon}_{\sigma \tau \kappa \omega}
            \\[0.5cm] \displaystyle
=({\delta}_{\mu \nu    }{\varepsilon}_{\alpha \beta \lambda \rho} -
  {\delta}_{\mu \alpha }{\varepsilon}_{\nu \beta \lambda \rho   } +
  {\delta}_{\mu \beta  }{\varepsilon}_{\nu \alpha \lambda \rho  } -
  {\delta}_{\mu \lambda}{\varepsilon}_{\nu \alpha \beta \rho    } +
  {\delta}_{\mu \rho   }{\varepsilon}_{\nu \alpha \beta \lambda })
{\varepsilon}_{\sigma \tau \kappa \omega}
            \\[0.5cm] \displaystyle
={\delta}_{\mu \nu }
\left| \matrix{
{\delta}_{\sigma \alpha}  & {\delta}_{\sigma \beta} &
{\delta}_{\sigma \lambda} & {\delta}_{\sigma \rho}  \\
{\delta}_{\tau \alpha}  & {\delta}_{\tau \beta} &
{\delta}_{\tau \lambda} & {\delta}_{\tau \rho}  \\
{\delta}_{\kappa \alpha}  & {\delta}_{\kappa \beta} &
{\delta}_{\kappa \lambda} & {\delta}_{\kappa \rho}  \\
{\delta}_{\omega \alpha}  & {\delta}_{\omega \beta} &
{\delta}_{\omega \lambda} & {\delta}_{\omega \rho}
} \right|
-{\delta}_{\mu \alpha }
\left| \matrix{
{\delta}_{\sigma \nu}     & {\delta}_{\sigma \beta} &
{\delta}_{\sigma \lambda} & {\delta}_{\sigma \rho}  \\
{\delta}_{\tau \nu}     & {\delta}_{\tau \beta} &
{\delta}_{\tau \lambda} & {\delta}_{\tau \rho}  \\
{\delta}_{\kappa \nu}     & {\delta}_{\kappa \beta} &
{\delta}_{\kappa \lambda} & {\delta}_{\kappa \rho}  \\
{\delta}_{\omega \nu}     & {\delta}_{\omega \beta} &
{\delta}_{\omega \lambda} & {\delta}_{\omega \rho}
} \right|  \\
            \\[0.5cm] \displaystyle
+ {\delta}_{\mu \beta}
\left| \matrix{
{\delta}_{\sigma \nu}     & {\delta}_{\sigma \alpha} &
{\delta}_{\sigma \lambda} & {\delta}_{\sigma \rho}   \\
{\delta}_{\tau \nu}     & {\delta}_{\tau \alpha} &
{\delta}_{\tau \lambda} & {\delta}_{\tau \rho}   \\
{\delta}_{\kappa \nu}     & {\delta}_{\kappa \alpha} &
{\delta}_{\kappa \lambda} & {\delta}_{\kappa \rho}   \\
{\delta}_{\omega \nu}     & {\delta}_{\omega \alpha} &
{\delta}_{\omega \lambda} & {\delta}_{\omega \rho}
} \right|
-{\delta}_{\mu \lambda}
\left| \matrix{
{\delta}_{\sigma \nu}   & {\delta}_{\sigma \alpha} &
{\delta}_{\sigma \beta} & {\delta}_{\sigma \rho}   \\
{\delta}_{\tau \nu}   & {\delta}_{\tau \alpha} &
{\delta}_{\tau \beta} & {\delta}_{\tau \rho}   \\
{\delta}_{\kappa \nu}   & {\delta}_{\kappa \alpha} &
{\delta}_{\kappa \beta} & {\delta}_{\kappa \rho}   \\
{\delta}_{\omega \nu}   & {\delta}_{\omega \alpha} &
{\delta}_{\omega \beta} & {\delta}_{\omega \rho}
} \right|   \\
            \\[0.5cm] \displaystyle
+ {\delta}_{\mu \rho}
\left| \matrix{
{\delta}_{\sigma \nu}   & {\delta}_{\sigma \alpha}  &
{\delta}_{\sigma \beta} & {\delta}_{\sigma \lambda} \\
{\delta}_{\tau \nu}   & {\delta}_{\tau \alpha}  &
{\delta}_{\tau \beta} & {\delta}_{\tau \lambda} \\
{\delta}_{\kappa \nu}   & {\delta}_{\kappa \alpha}  &
{\delta}_{\kappa \beta} & {\delta}_{\kappa \lambda} \\
{\delta}_{\omega \nu}   & {\delta}_{\omega \alpha}  &
{\delta}_{\omega \beta} & {\delta}_{\omega \lambda}
} \right| =
\left| \matrix{
{\delta}_{\mu \nu}   & {\delta}_{\mu \alpha}  &
{\delta}_{\mu \beta} & {\delta}_{\mu \lambda} &
{\delta}_{\mu \rho}  \\
{\delta}_{\sigma \nu}   & {\delta}_{\sigma \alpha}  &
{\delta}_{\sigma \beta} & {\delta}_{\sigma \lambda} &
{\delta}_{\sigma \rho}  \\
{\delta}_{\tau \nu}   & {\delta}_{\tau \alpha}  &
{\delta}_{\tau \beta} & {\delta}_{\tau \lambda} &
{\delta}_{\tau \rho}  \\
{\delta}_{\kappa \nu}   & {\delta}_{\kappa \alpha}  &
{\delta}_{\kappa \beta} & {\delta}_{\kappa \lambda} &
{\delta}_{\kappa \rho}  \\
{\delta}_{\omega \nu}   & {\delta}_{\omega \alpha}  &
{\delta}_{\omega \beta} & {\delta}_{\omega \lambda} &
{\delta}_{\omega \rho}
} \right| \equiv 0 \;\; .
\end {array}
\label{e2}
\end {equation}

By analogy with the Gram determinant, we introduce the notation
(see \cite{r1})
\begin {equation}
\begin {array}{l} \displaystyle
\left| \matrix{
{\delta}_{\mu \nu}   & {\delta}_{\mu \alpha}  &
{\delta}_{\mu \beta} & {\delta}_{\mu \lambda} &
{\delta}_{\mu \rho}  \\
{\delta}_{\sigma \nu}   & {\delta}_{\sigma \alpha}  &
{\delta}_{\sigma \beta} & {\delta}_{\sigma \lambda} &
{\delta}_{\sigma \rho}  \\
{\delta}_{\tau \nu}   & {\delta}_{\tau \alpha}  &
{\delta}_{\tau \beta} & {\delta}_{\tau \lambda} &
{\delta}_{\tau \rho}  \\
{\delta}_{\kappa \nu}   & {\delta}_{\kappa \alpha}  &
{\delta}_{\kappa \beta} & {\delta}_{\kappa \lambda} &
{\delta}_{\kappa \rho}  \\
{\delta}_{\omega \nu}   & {\delta}_{\omega \alpha}  &
{\delta}_{\omega \beta} & {\delta}_{\omega \lambda} &
{\delta}_{\omega \rho}
} \right| =
G\pmatrix{\mu & \sigma & \tau  & \kappa  & \omega \\
          \nu & \alpha & \beta & \lambda & \rho    }  \;\; .
\end {array}
\label{e3}
\end {equation}

We consider some consequences of the identity (\ref{e2}).
\begin{enumerate}
\item
It follows from the properties of determinants that
\begin {equation}
\begin {array}{l} \displaystyle
G\pmatrix{\mu & \sigma & \tau  & \kappa  & \omega & \chi \\
          \nu & \alpha & \beta & \lambda & \rho   & \varphi}
\equiv 0 \;\; ,
\end {array}
\label{e4}
\end {equation}

\begin {equation}
\begin {array}{l} \displaystyle
G\pmatrix{\mu & \sigma & \tau  & \kappa  & \omega & \chi    & \xi \\
          \nu & \alpha & \beta & \lambda & \rho   & \varphi & \pi}
          \equiv 0
\end {array}
\label{e5}
\end {equation}
etc. We note that the identity (\ref{e2}) is valid only in a space
of dimension not higher than 4, the identity (\ref{e4}) is valid
in a space of dimension not higher than 5, etc.

\item
We contract (\ref{e2}) with 10 arbitrary 4-vectors:
\begin {equation}
\begin {array}{l} \displaystyle
     G\pmatrix{\mu & \sigma & \tau  & \kappa  & \omega \\
               \nu & \alpha & \beta & \lambda & \rho}
     a_{\mu} b_{\sigma} c_{\tau} d_{\kappa }
     e_{\omega} f_{\nu} g_{\alpha} m_{\beta} n_{\lambda} p_{\rho}
     = G\pmatrix{a & b & c & d & e \\
                 f & g & m & n & p} \equiv 0 \;\; .
\end {array}
\label{e6}
\end {equation}

The identity (\ref{e6}) reflects the well-known fact that a Gram
determinant of fifth order and higher in Minkowski space is always
equal to 0.

\item
We consider
\begin {equation}
\begin {array}{l} \displaystyle
    G\pmatrix{\mu & \sigma & \tau  & \kappa  & \omega \\
              \nu & \alpha & \beta & \lambda & \rho }
    (a_1)_{\mu} (a_2)_{\sigma} (a_3)_{\tau} (a_4)_{\kappa}
    (a_5)_{\omega} (a_1)_{\nu} (a_2)_{\alpha} (a_3)_{\beta}
    (a_4)_{\lambda}  \\
            \\[0.5cm] \displaystyle
= G\pmatrix{a_1 & a_2 & a_3 & a_4 & a_5 \\
            a_1 & a_2 & a_3 & a_4 & \rho}
= (a_1)_{\rho} G\pmatrix{a_2 & a_3 & a_4 & a_5 \\
                         a_1 & a_2 & a_3 & a_4 }      \\
            \\[0.5cm] \displaystyle
- (a_2)_{\rho} G\pmatrix{a_1 & a_3 & a_4 & a_5 \\
                         a_1 & a_2 & a_3 & a_4 }
+ (a_3)_{\rho} G\pmatrix{a_1 & a_2 & a_4 & a_5 \\
                         a_1 & a_2 & a_3 & a_4 }        \\
            \\[0.5cm] \displaystyle
- (a_4)_{\rho} G\pmatrix{a_1 & a_2 & a_3 & a_5 \\
                         a_1 & a_2 & a_3 & a_4 }
+ (a_5)_{\rho} G\pmatrix{a_1 & a_2 & a_3 & a_4 \\
                         a_1 & a_2 & a_3 & a_4 } \equiv 0 \;\; .
\end {array}
\label{e7}
\end {equation}
The identity (\ref{e7}) can be used to decompose an arbitrary
4-vector $a_5$ with respect to the basis consisting of the
4-vectors $a_1$, $a_2$, $a_3$, $a_4$.

\item
Identities of the type (\ref{e2}), (\ref{e4}), (\ref{e5}) and
others like them can be used to simplify the expressions for the
traces of a product of 10 and more Dirac ${\gamma}$ matrices. For
example, the expression for
$$
\displaystyle
Tr[ {\hat a}_1 {\hat a}_2  {\hat a}_3
    {\hat a}_4 {\gamma}_{\mu}
    {\hat a}_1 {\hat a}_2  {\hat a}_3
    {\hat a}_4 {\gamma}_{\nu} ]
$$
($a_1$, $a_2$, $a_3$, $a_4$
  are arbitrary 4-vectors,
$
{\hat a} = a_{\rho} {\gamma}_{\rho}
$), calculated by means of the computer system REDUCE,
contains 100 terms.

However, the expression\footnotemark
\footnotetext[2]
{
$$
\displaystyle
G\pmatrix{a_1  & a_2  & a_3 & a_4 & \mu  \cr
          a_1  & a_2  & a_3 & a_4 & \nu  }
=  G\pmatrix{\sigma  & \tau  & \kappa   & \omega & \mu  \cr
             \alpha  & \beta & \lambda  & \rho   & \nu  }
(a_1)_{\sigma} (a_2)_{\tau}  (a_3)_{\kappa}  (a_4)_{\omega}
(a_1)_{\alpha} (a_2)_{\beta} (a_3)_{\lambda} (a_4)_{\rho }
\, .
$$
}
\begin {equation}
\begin {array}{c} \displaystyle
Tr[ {\hat a}_1 {\hat a}_2  {\hat a}_3
    {\hat a}_4 {\gamma}_{\mu}
    {\hat a}_1 {\hat a}_2  {\hat a}_3
    {\hat a}_4 {\gamma}_{\nu} ]
-  16  G\pmatrix{a_1 & a_2 & a_3 & a_4 & \mu \cr
                      a_1 & a_2 & a_3 & a_4 & \nu  }
            \\[0.5cm] \displaystyle
= 8 \{ (a_1)_{\mu} (a_1)_{\nu} (a_2)^2 (a_3)^2 (a_4)^2
- [ (a_1)_{\mu} (a_2)_{\nu} + (a_1)_{\nu} (a_2)_{\mu}]
    (a_3)^2 (a_4)^2 (a_1 a_2)
            \\[0.5cm] \displaystyle
+ [(a_1)_{\mu} (a_3)_{\nu} + (a_1)_{\nu} (a_3)_{\mu}]
   (a_4)^2 [2 (a_1 a_2) (a_2 a_3) - (a_2)^2 (a_1 a_3)]
            \\[0.5cm] \displaystyle
+ [(a_1)_{\mu} (a_4)_{\nu} + (a_1)_{\nu} (a_4)_{\mu} ]
[-4 (a_1 a_2) (a_2 a_3) (a_3 a_4)
+ 2(a_3)^2 (a_1 a_2) (a_2 a_4)
            \\[0.5cm] \displaystyle
+ 2 (a_2)^2 (a_1 a_3) (a_3 a_4)
- (a_2)^2 (a_3)^2 (a_1 a_4) ]
            \\[0.5cm] \displaystyle
+ (a_2)_{\mu} (a_2)_{\nu} (a_1)^2 (a_3)^2 (a_4)^2
- [ (a_2)_{\mu} (a_3)_{\nu} + (a_2)_{\nu} (a_3)_{\mu} ]
    (a_1)^2 (a_4)^2 (a_2 a_3)
            \\[0.5cm] \displaystyle
+ [ (a_2)_{\mu} (a_4)_{\nu} + (a_2)_{\nu} (a_4)_{\mu} ]
(a_1)^2  [2 (a_2 a_3) (a_3 a_4) - (a_3)^2 (a_2 a_4) ]
            \\[0.5cm] \displaystyle
+ (a_3)_{\mu} (a_3)_{\nu} (a_1)^2 (a_2)^2 (a_4)^2
- [ (a_3)_{\mu} (a_4)_{\nu} + (a_3)_{\nu} (a_4)_{\mu} ]
    (a_1)^2 (a_2)^2 (a_3 a_4)
            \\[0.5cm] \displaystyle
+ (a_4)_{\mu} (a_4)_{\nu} (a_1)^2 (a_2)^2 (a_3)^2
            \\[0.5cm] \displaystyle
+ \delta_{\mu \nu} [ 4 (a_1 a_2) (a_2 a_3) (a_3 a_4)
  (a_1 a_4) - 2 (a_1)^2 (a_2 a_3) (a_3 a_4) (a_2 a_4)
            \\[0.5cm] \displaystyle
- 2 (a_2)^2 (a_1 a_3) (a_3 a_4) (a_1 a_4)
- 2 (a_3)^2 (a_1 a_2) (a_2 a_4) (a_1 a_4)
            \\[0.5cm] \displaystyle
- 2 (a_4)^2 (a_1 a_2) (a_2 a_3) (a_1 a_3) -
 {3 \over 2} (a_1)^2 (a_2)^2 (a_3)^2 (a_4)^2
            \\[0.5cm] \displaystyle
+ (a_1)^2 (a_2)^2 (a_3 a_4)^2 + (a_1)^2 (a_3)^2
  (a_2 a_4)^2 + (a_1)^2 (a_4)^2 (a_2 a_3)^2
            \\[0.5cm] \displaystyle
+ (a_2)^2 (a_3)^2 (a_1 a_4)^2 + (a_2)^2 (a_4)^2
  (a_1 a_3)^2 + (a_3)^2 (a_4)^2 (a_1 a_2)^2 ] \}
\, ,
\end {array}
\label{e8}
\end {equation}
which is identical to it, contains only 38 terms.
\end{enumerate}

We note that the same result can also be obtained without direct
use of the identity (\ref{e2}). For this, it is necessary to use
the expression (see the Appendix)
\begin {equation}
\displaystyle
    {\hat a}_1  \ldots {\hat a}_{2n}
= - {\hat a}_{2n} \ldots {\hat a}_1
+ {1 \over 2}  Tr [ {\hat a}_1 \ldots {\hat a}_{2n} ]
+ {1 \over 2} {\gamma}_5
Tr [ {\gamma}_5 {\hat a}_1 \ldots {\hat a}_{2n} ] \;\; .
\label{e9}
\end {equation}

We apply (\ref{e9}) twice:
$$
\begin {array}{c} \displaystyle
Tr [ {\hat a}_1 {\hat a}_2 {\hat a}_3 {\hat a}_4
     {\gamma}_{\mu}
     {\hat a}_1 {\hat a}_2 {\hat a}_3 {\hat a}_4
     {\gamma}_{\nu} ] =
- Tr [ {\hat a}_4 {\hat a}_3 {\hat a}_2 {\hat a}_1
       {\gamma}_{\mu}
       {\hat a}_1 {\hat a}_2 {\hat a}_3 {\hat a}_4
       {\gamma}_{\nu} ]
            \\[0.5cm] \displaystyle
+ {1 \over 2}
       Tr [ {\hat a}_1 {\hat a}_2 {\hat a}_3 {\hat a}_4 ]
  Tr [ {\gamma}_{\mu}
            {\hat a}_1 {\hat a}_2 {\hat a}_3 {\hat a}_4
            {\gamma}_{\nu} ]
+ {1 \over 2}
  Tr [ {\gamma}_5
            {\hat a}_1 {\hat a}_2 {\hat a}_3 {\hat a}_4 ]
  Tr [ {\gamma}_5 {\gamma}_{\mu}
            {\hat a}_1 {\hat a}_2 {\hat a}_3 {\hat a}_4
            {\gamma}_{\nu} ]
              \\[0.5cm] \displaystyle
= - Tr [ {\hat a}_1 {\hat a}_2 {\hat a}_3 {\hat a}_4
         {\gamma}_{\mu}
         {\hat a}_4 {\hat a}_3 {\hat a}_2 {\hat a}_1
         {\gamma}_{\nu} ]
              \\[0.5cm] \displaystyle
+ {1 \over 2}
       Tr [ {\hat a}_1 {\hat a}_2 {\hat a}_3 {\hat a}_4 ]
  Tr [ {\hat a}_1 {\hat a}_2 {\hat a}_3 {\hat a}_4
            {\gamma}_{\mu} {\gamma}_{\nu} ]
+ {1 \over 2}
  Tr [ {\gamma}_5
            {\hat a}_1 {\hat a}_2 {\hat a}_3 {\hat a}_4 ]
  Tr [ {\hat a}_1 {\hat a}_2 {\hat a}_3 {\hat a}_4
            {\gamma}_{\mu} {\gamma}_5 {\gamma}_{\nu} ]
              \\[0.5cm] \displaystyle
= - {1 \over 2} \left(
  Tr [ {\hat a}_4 {\hat a}_3 {\hat a}_2 {\hat a}_1
       {\gamma}_{\mu}
       {\hat a}_1 {\hat a}_2 {\hat a}_3 {\hat a}_4
       {\gamma}_{\nu} ]
+ Tr [ {\hat a}_4 {\hat a}_3 {\hat a}_2 {\hat a}_1
         {\gamma}_{\nu}
       {\hat a}_1 {\hat a}_2 {\hat a}_3 {\hat a}_4
         {\gamma}_{\mu} ] \right)
              \\[0.5cm] \displaystyle
+ {1 \over 4} Tr [ {\hat a}_1 {\hat a}_2 {\hat a}_3 {\hat a}_4 ]
  Tr [ {\hat a}_1 {\hat a}_2 {\hat a}_3 {\hat a}_4
         ( {\gamma}_{\nu} {\gamma}_{\mu}
          + {\gamma}_{\mu} {\gamma}_{\nu} ) ]
              \\[0.5cm] \displaystyle
+ {1 \over 4}
  Tr [ {\gamma}_5 {\hat a}_1 {\hat a}_2 {\hat a}_3 {\hat a}_4 ]
  Tr [ {\gamma}_5
    (- {\gamma}_{\nu} {\gamma}_{\mu}
     - {\gamma}_{\mu} {\gamma}_{\nu} )
       {\hat a}_1 {\hat a}_2 {\hat a}_3 {\hat a}_4 ]
              \\[0.5cm] \displaystyle
= - {1 \over 2} \left(
  Tr [ {\hat a}_4 {\hat a}_3 {\hat a}_2 {\hat a}_1 {\gamma}_{\mu}
       {\hat a}_1 {\hat a}_2 {\hat a}_3 {\hat a}_4 {\gamma}_{\nu} ]
+ Tr [ {\hat a}_4 {\hat a}_3 {\hat a}_2 {\hat a}_1 {\gamma}_{\nu}
       {\hat a}_1 {\hat a}_2 {\hat a}_3 {\hat a}_4 {\gamma}_{\mu} ]
       \right)
              \\[0.5cm] \displaystyle
+ {1 \over 2} {\delta}_{\mu \nu}
\{ (Tr [ {\hat a}_1 {\hat a}_2 {\hat a}_3 {\hat a}_4 ] )^2
- (Tr [ {\gamma}_5
  {\hat a}_1 {\hat a}_2 {\hat a}_3 {\hat a}_4 ] )^2 \}
\;\; .
\end {array}
$$
Further calculations give (\ref{e8}).
          \newline

{\bf APPENDIX}

We write down one of the formulas of Fierz transformations
(see~\cite{r2}):
\begin {equation}
\displaystyle
{1 \over 2} { \left[ (1 \pm {\gamma}_5) {\gamma}_{\mu} \right] }_{ij}
            { \left[ (1 \mp {\gamma}_5 ) {\gamma}_{\mu} \right] }_{kl}
= { (1 \pm {\gamma}_5) }_{il} { (1 \mp {\gamma}_5) }_{kj}
\label{e10}
\end {equation}
($i$, $j$, $k$, $l$ are indices that label the components of
$4 \times 4$ matrices).

A consequence of (\ref{e10}) is the formula
\begin {equation}
\begin {array}{c} \displaystyle
{1 \over 2} { \left[ (1 \pm {\gamma}_5) {\gamma}_{\mu} \right]
}_{ij}
    Q_{jk} { \left[ (1 \mp {\gamma}_5) {\gamma}_{\mu} \right] }_{kl}
=  { (1 \pm {\gamma}_5) }_{il} { (1 \mp {\gamma}_5) }_{kj} Q_{jk}
            \\[0.5cm] \displaystyle
= { (1 \pm  {\gamma}_5) }_{il} Tr [ (1 \mp {\gamma}_5) Q ]
\end {array}
\label{e11}
\end {equation}
(here, $Q$ is an arbitrary $4 \times 4$ matrix).

In addition, we have (see \cite{r2})
\begin {equation}
\displaystyle
{\gamma}_{\mu}
  {\hat a}_1 \ldots {\hat a}_{2n-1} {\hat a}_{2n}
{\gamma}_{\mu}
= 2 ( {\hat a}_{2n} {\hat a}_1 \ldots {\hat a}_{2n-1}
    + {\hat a}_{2n-1} \ldots {\hat a}_1 {\hat a}_{2n} )
\label{e12}
\end {equation}
(here, $a_1, \ldots, a_{2n}$ are arbitrary 4-vectors).
Now suppose that in (\ref{e11})
$$
\displaystyle
Q = {\hat a}_2 \ldots {\hat a}_{2n} {\hat a}_1 \;\; .
$$

With allowance for (\ref{e11}) and (\ref{e12}), we have
\begin {equation}
\begin {array}{c} \displaystyle
{1 \over  2} (1 \pm {\gamma}_5) {\gamma}_{\mu}
 {\hat a}_2 \ldots {\hat a}_{2n} {\hat a}_1
(1 \mp {\gamma}_5) {\gamma}_{\mu}
=  (1 \pm {\gamma}_5) Tr[ (1 \mp {\gamma}_5)
 {\hat a}_2 \ldots {\hat a}_{2n} {\hat a}_1 ]
            \\[0.5cm] \displaystyle
=  (1 \pm {\gamma}_5) Tr[ (1 \pm {\gamma}_5)
                             {\hat a}_1 \ldots {\hat a}_{2n} ]
= (1 \pm {\gamma}_5) {\gamma}_{\mu}
   {\hat a}_2 \ldots {\hat a}_{2n} {\hat a}_1
   {\gamma}_{\mu}
            \\[0.5cm] \displaystyle
=  2 (1 \pm {\gamma}_5 ) ( {\hat a}_1 \ldots {\hat a}_{2n}
                      + {\hat a}_{2n} \ldots {\hat a}_1 ) \;\; .
\end {array}
\label{e13}
\end {equation}

Thus
\begin {equation}
\displaystyle
(1 \pm {\gamma}_5) ( {\hat a}_1 \ldots {\hat a}_{2n}
                    + {\hat a}_{2n} \ldots {\hat a}_1 )
= {1\over  2} (1 \pm {\gamma}_5)
Tr [ (1 \pm {\gamma}_5) {\hat a}_1 \ldots {\hat a}_{2n} ]
\label{e14}
\end {equation}
for all 4-vectors  $a_1, \ldots, a_{2n}$ .

The expression (\ref{e9}) is a direct consequence of (\ref{e14}).
Note that the expressions (\ref{e9}) and (\ref{e14}) generalize
the corresponding expressions obtained in  \cite{r3}, \cite{r4}.

\end {document}